\documentclass[twocolumn,showpacs,preprintnumbers,amsmath,amssymb]{revtex4}

\def\be{\begin{equation}}
\def\ee{\end{equation}}
\def\bea{\begin{eqnarray}}
\def\eea{\end{eqnarray}}

\def\m{\mu}
\def\n{\nu}

\def\p{\partial}
\def\a{\alpha}
\def\b{\beta}
\def\t{\theta}

\usepackage{graphicx}
\usepackage{dcolumn}
\usepackage{bm}
\usepackage{amsmath}
\usepackage{amssymb}
\usepackage{epsfig}
\usepackage{color}

\begin{document}

\title{Tensor of the energy - momentum and forbiddance of the classical Lump}
\author{Abolfazl\ Jafari}
\affiliation{Department of Physics, Faculty of Science,
Shahrekord University, P. O. Box 115, Shahrekord, Iran.\\
jafari-ab@sci.sku.ac.ir}
\date{\today }

\begin{abstract}
\textbf{Abstract}
We show that the static solutions of massive classical fields are forbidden in
the Minkowski space time with dimensions higher than $2$.
The purpose of this study is to generalize the forbiddance of the time independent
localized classical fields in general space time.
We generalize the Derrick theorem as unstable stationary localized solutions for the nonlinear
wave equation or the nonlinear Klein - Gordon equation in Minkowski space time.
Our approach employs the tensor of the energy - momentum based
on Weber's method. 
We exploite the Bateman - Caldilora - Kanai model and the Deser method for the research of the Yang - Mills Lumps as massive solutions.
We also use Riemann's coordinates for parametrization of tangent space as a local framework.
The important point of this article is the homomorphism between the Minkowsian and pseudo Riemann's manifolds.
\end{abstract}


\maketitle

\section{Introduction and Conclusions}

Derrick's theorem is an argument proposed by a physicist G.H. Derrick, who has shown the
instability of
stationary localized solutions for a nonlinear
wave equation or the nonlinear Klein - Gordon equation in dimensions three and higher\cite{derrick,kolokolov}.
Also, Coleman and Deser have published the cardinal theorem about
the absence of time independent solutions for the source free Yang - Mills equation:
There are no static solutions for the source free Yang- Mills field's equation of motion\cite{coleman1, deser}.
Instability of the static solutions in particular is satisfactory in Chromodynamics.
This can be generalized to include quantum electrodynamics entirely.
Perhaps, one of the important consequences of this lemma is the absence of classical glueballs.
That is, the Classical glueballs can not occur in the space time with $3+1$-dimensions.

There are some kinds of bound states which are named the glueballs of QCD.
The glueballs as colorless bound states of the massive composed
exclusively of the gauge field quanta (gluons) which gets the mass in the Higgs mechanism.
The quantum electrodynamics as a massless theorem does not consist of self - interaction,
therefore we do not encounter the bound states\cite{jafari}.

The name of the static and localized field is Lump.
The Lump is one of the solutions of the equation of the motion of the field and is localized at the space time.
The time dependent Lumps as dynamic Lumps radiate energy but, static Lumps can not radiate.
According to special relativity, absence of the Lumps related to the massive or massless fields have different consequences.
Since, the Lorentz transformation can be always found for the massive fields, the absence of the massive
Lumps can be generalized to the rest frame.

According to Coleman and Deser, there are no finite-energy non-singular solutions of
the classical Yang-Mills theory in four-dimensional Minkowski space time that do not
radiate energy out to the spatial infinity.
In a linear classical field theory, like free electrodynamics, any non-singular initial
configuration of fields of finite energy will eventually spread out over all space;
whatever the initial configuration, the final state is simply outgoing radiation.
In
contrast, certain non-linear field theories are known to have finite-energy nonsingular
solutions that can be described as Lumps of energy held together by their
own self-interaction, in which no energy is radiated to spatial infinity.
Also in Refs. \cite{coleman1, coleman2} the authors have been shown that if the static solutions in quantum field theory are forbidden,
they will be forbidden in the classical limit.
That is the nonexistence of quantum
glueballs says that nothing against the existence of classical glueballs\cite{coleman1, coleman2, deser}.
Furthermore, we always find tangent spaces in which case an approximated version of physics can be extracted.
There can always be found one way to study quantum field theory on any asymptotic universe.

Our study of field theories in an asymptotic universe is based on a given pseudo Riemann's manifolds.
There is one class of approximated quantum field theory which it is different from usual methods.
In the present, we show that a new definition of approximated version of the quantum field theory is available for a limited region
\cite{speliotopoulos, saha}.
The new method applies only and only if we accept the three following principles:
all changes will occur in metric and its related tensors. 
The proper time equals usual time at locally and 
there can always be found an equality between proper time and usual time at locally.
Our purpose from importing the tangent space is that if we could prove the absence of the Lumps locally, in that case
the obtained theorem will be developed to the global universe.

There is a famous issue in agreement with the Derrick's theorem:
the existence of the Sine - Gordon equation as a nonlinear hyperbolic partial differential equation in $1+1$ space time dimensions. 
It was originally introduced by Edmond Bour (1862) in the course of study of surfaces of constant negative 
curvature as the Gauss - Codazzi equation for surfaces of curvature, $-1$ in $3$-space, and rediscovered by Frenkel and Kontorova (1939) in their 
study of crystal dislocations. 
This equation attracted a lot of attention in the 1970s due to the presence of soliton solutions.
The Sine - Gordon equation is the Euler - Lagrange equation of the field whose Lagrangian density is given by
\begin{align}
\mathfrak{L}_\mathrm{SG}=\mathfrak{L}_\mathrm{KG}+\cos{\phi}.
\end{align}
An interesting feature of the Sine - Gordon equation is the existence of soliton and multisoliton solutions.
The Sine - Gordon equation has many soliton solutions
which can be obtained through continued application of the B$\ddot{a}$cklund transform to the other soliton solution
\cite{raja,dod}.
According to the general relativity principles,
one can never find coordinate systems in curved space time with $\Gamma^{\delta}_{\a\b}=0$ everywhere.
But one can always construct local initial frames at a given event and as viewed in such frames,
free particles must move along straight lines, at least locally.
It means that $\Gamma^{\delta}_{\a\b}$
must vanishes up to the first order of the Riemannian curvature tensors which implies that the tangent
space related to the pseudo Riemann universes will be flat while it was filled with gravitational effects.

The Einstein equivalence principle postulates 
that in the presence of a gravitational field, special relativity will be available in a
non relativistic regimes.
This implies that in the general relativity, one can find a locally constructed inertial frame.
Indeed, a local inertial frame can be found for each given point ($\rho_0$) interior of the space time
where the components of the metric tensor satisfies the following equation
\begin{align}
g_{\a\b,\m}(\rho_0)=0.
\end{align}
The coordinates of such a frame are called the \textit{Riemann's coordinates}\cite{nestrov,misner,weber,maggiore,inverno}.
A specialization of the Riemann's coordinates is done when the coordinates line going through the event is taken as geodesics.
Such coordinates are
called \textit{Riemann's normal coordinates}.
In normal coordinates each spacelike hypersurface of constant $x^0$ is normal to the geodesic and spatial coordinates
$x^i$ of a point occurring at time $x^0$. $x^i$ can be taken as the Riemann's normal coordinates.
The metric components have the following forms in the Riemann's coordinates up to the first order of Riemann's tensors
\begin{align}
\label{metric}
g_{00}=-1-R_{0l0m}x^lx^m,
g_{0i}=g^{0i}=-\frac{2}{3}R_{0lim}x^lx^m,
\cr
g_{ij}=\delta_{ij}-\frac{1}{3}R_{iljm}x^lx^m,
g^{00}=-1+R_{0l0m}x^lx^m,
\cr
g^{ij}=\delta^{ij}+\frac{1}{3}R^{ij}_{lm} x^lx^m,
g=-1+\frac{1}{3}(R_{lm}-2R_{0l0m})x^lx^m,
\end{align}
and the affine connections (Christoffel multipliers) are
\begin{align}
\label{affine}
\Gamma^{0}_{00}=0,\ \ \ \Gamma^{0}_{ij}=\frac{1}{3}(R_{0ijm}+R_{0jim})x^m,\cr
\Gamma^{0}_{0i}=R_{0i0m}x^m,\ \ \ \Gamma^{i}_{jk}=\frac{1}{3}(R_{jikm}+R_{kijm})x^m,\cr
\Gamma^{i}_{0j}=R_{0mji}x^m,\ \ \ \Gamma^{i}_{00}=R_{0i0m}x^m,
\end{align}
where Cartesian indices run from 1 to 3 and the constructed tangent space is similar to the Minkowski space time.
We mention that the raising of the Lorentz indices indicated on the tangent space done with 
Eqs.(\ref{metric}),
due to
ignoring higher powers of the Riemann parameters.
Also, the tangent space is constructed at a constant Riemann's curvature tensors\cite{parker}.

The common tool for all approximated methods is working with local coordinates.
The validity of above sentences is limited to non - covariant observer,
but we show that the obtained result can be generalized to consists of the global universe.
In this manner, the research of the Lumps are different for the static and dynamic universe
such as the static Schwarzschild and the presence of gravitational waves.

\section{Presentation of the theory}
This work is based on the Derrick theorem using the Coleman and Deser methods.
We consider space time to have one time dimension and $(d-1)$ space dimensions.
The vanishing of the energy - momentum tensor for the static systems excludes finite energy time - independent solutions of the source - free
theories.
Because the scalar current is traceless ($\int\ d^{d-1}x\ T^{\m}_{\m}=0$) and this can be extracted directly from the equation of the motion.
This implies that the static solutions are forbidden.

At a large distance from the center of the energy of the Lump, the strength's field
must goes to zero, because we consider the localized fields.
This condition is necessary for the definition of concentrated fields.
By introducing the localized and real scalar fields $\phi$ and non - abelian strength tensor $F_a^{\m\n}=f_a^{\m\n}-\imath g[A_\m,A_\n]_a$ with abelian form of the strength tensor of fields $f^a_{\m\n}=\p_\m A^a_\n-\p_\n A^a_\m$,
we have the beginning of the proof.
If the metric of $d$-dimensional space time is similar to that of Minkowski so
the Lagrangian which describes the theories of classical field (scalar field, Proca and Gluons) becomes
\begin{align}\label{intial lagrange}
\mathfrak{L}_{\mathrm{SFT}}=-\frac{1}{2}(\eta^{\m\n}\p_\m\phi\ \p_\n\phi+m^2\phi^2)+V(\phi),
\cr
\mathfrak{L}_{\mathrm{Proca}}=-\frac{1}{4}f^{\m\n} f_{\m\n}+\frac{1}{2}m^2A_\m A^\m,
\cr
\mathfrak{L}_{\mathrm{Gluons}}=Tr(-\frac{1}{4}F^{\m\n} F_{\m\n})
\cr=-\frac{1}{4}f^{\m\n}_a f^a_{\m\n}+2gc^{a}_{bc}f^{\m\n}_aA^b_\m A^c_\n+
g^2[A^\m,A^\n]_a[A_\m,A_\n]^a,
\end{align}
so that the potential function satisfies the condition $V(\phi=0)=0$.
Here, $\mathfrak{L}$ stands for the relevant Lagrangian density, all Greek indices run from 0 to 3 and $m$ is the field mass.
Latin indices indicate the relevant generators of the symmetry group and $C^a_{bc}$ is the constant of the structure of the Lie algebra.
Common electrodynamics theory is described with a second case in Eq.(\ref{intial lagrange}) when $m=0$.
By variation of the action with respect to the tensor of metric\cite{parker1, kleinert1, kleinert2}
the relevant tensors of the energy - momentum 
become
\begin{align}
\label{energy - momentum}
T^{\m}_{\n\ {\mathrm{SFT}}}=-\eta^{\m}_{\n}\mathfrak{L}_{\mathrm{SFT}}-\p^\m\phi\ \p_\n\phi,
\cr
T^{\m}_{\n\ {\mathrm{Proca}}}=-\eta^{\m}_{\n}\mathfrak{L}_{\mathrm{Proca}}-\p^\m A^\kappa\ \p_\n A_\kappa
\cr
=-f^{\m\a}f_{\a\n}-\frac{1}{4}\eta^\m_\n f^{\a\b}f_{\a\b}+\frac{1}{2}m^2\eta^\m_\n A^\a A_\a,
\cr
T^{\m}_{\n\ {\mathrm{Gluons}}}=-F^{\m\a}_{a}F^{a}_{\a\n}-\frac{1}{4}\eta^{\m}_{\n}F^{\a\b}F_{\a\b}.
\end{align}
By applying the least principle, the equations of the motion of the fields are given by
\begin{align}
\p_\m\p^\m\phi-m^2\phi+\frac{d}{d\phi}V(\phi)=0,
\cr
\p_\m f^{\m\n}+m^2 A^\n=0,
\cr
\p_\m f^{\m\n}_a+gc_{ab}^{c}A^b_\m f^{\m\n}_c=0.
\end{align}
In the common form, the tensor of the energy - momentum obeys the following:
\begin{align}
\p_\m T^{\m\n}=0.
\end{align}
The equations of motion of the fields and the energy - momentum tensors always are two linearly independent expect for
$1+1$ space time dimensions.
In a locally inertial rest frame, with $g_{\m\n}$ diagonal and time independent fields, the matrix elements of the energy - momentum tensors are
\begin{align}
T^{\m\n}=\left(
\begin{array}{cc}
T^{00} & 0 \\ 
0 & T^{11}
\end{array}
\right),
\end{align}
where $\p_0 T^{0\n}=0$. 
By these tools,
the equation of the motion of the energy - momentum tensor for the scalar field gets the following
\begin{align} 
\p_1 T^{11}_{\mathrm{SFT}}=-\p^1\mathfrak{L}-\p_1(\p^1\phi\ \p^1\phi)
=
-\p^1(-\frac{1}{2}(\p^1\phi\p_1\phi
\cr
+m^2\phi^2)
+V(\phi))-\p_1(\p^1\phi\ \p^1\phi)
=
(-\p^1\p_1\phi+m^2\phi
\cr
-\frac{d}{d\phi}V(\phi))\p^1\phi.
\end{align}
Similar calculations for the case of Proca can also be obtained. 
That is,
\begin{align} 
\p_1 T^{11}_{\mathrm{Proca}}=-\p_1(-f^{10}f_{01}-\frac{1}{4}(2f^{10}f_{10})+\frac{1}{2}m^2A_\a A^\a)
\cr
=f_{10}\p_1 f^{10}+m^2 A_\a \p_1 A^\a.
\end{align}
One makes the assumption that the classical field is time independent and by taking Lorentz gauge into account, we can write, $A_\a\p_1 A^\a=A_0\p_1 A^0+A_1\p_1 A^1=A_0\p_1 A^0$, 
scince $f_{10}=\p_1 A_0-\p_0 A_1=\p_0 A_1$, so we get to
\begin{align} 
\p_1 T^{11}_{\mathrm{Proca}}=\p_1 A_0(\p_1 f^{10}+m^2 A^0),
\end{align}
also, for Gluons,
\begin{align}
\p_1T^{1}_{1\ {\mathrm{Gluons}}}=\p_1(-F_{10}^a F_a^{01}-\frac{1}{4}\eta^1_1(2F^a_{10}F_a^{10}))
\cr
=\frac{1}{2}\p_1(F^a_{10}F_a^{10})=(\p_1F^{10}_a) F^a_{10},
\end{align}
but, from the equation of the motion of the Gluons we have,
\begin{align}
\p_1 F^{10}_a=-gc^{bc}_a A_b^1 F_c^{10},
\end{align}
where,
\begin{align}
(\p_1 F_a^{10})F^a_{10}=-gc_{a}^{bc}A^1_b F_c^{10}F^a_{10}.
\end{align}
All of the above vanish for the equation of the motion of the classical fields.
So far, the equations of the motion of the energy - momentum tensors 
did not give us anything extra
to the equation of the motion of the fields, because
they are not all independent. 
So for $1+1$ space time dimensions, the free scalar Lump and another classical Lumps can occur and we learn nothing further from the energy - momentum tensor viewpoint. 
Obtained results have an argument with exsistance of the classical Lumps.
So, the classical glueballs as a Lump of Gluons, happens in $1+1$ space time dimensions.
This gives us an additional dimension to the result of the Ref \cite{coleman}. 
For the dimensions higher than $1+1$, in Ref. \cite{coleman} aothurs have been investigate existing condition of the Lump for the Yang - Mills theory.
From Ref. \cite{coleman} the classical Lump of the Yang - Mills can not occur in $3+1$ dimensions.

Now, exclusively for free scalar fields, when $V(\phi)=0$, the Derrick's theorem says:
"the solution for the scalar fields are forbidden except in two dimensional space time."
In confirmation of the Derrick' theorem and based on the behavior of the energy - momentum tensor,
we can suggest a lemma:\\
\textbf{Lemma}: \textit{The Lump of the scalar fields is forbidden in certain space time expect $1+1$ dimensions}.

\textbf{Proof}:
The substantiation of Derrick's theorem is based on the scaling process and profited from the Klein - Gordon
equation without taking account of the dynamic of the energy - momentum tensor,
but, we take into account the dynamic of the energy - momentum tensor.
During to the equation of the motion for the energy - momentum tensor and due to the zero values for the Lagrangian, $\mathfrak{L}_{\mathrm{SFT}}$,
on the mass shell, it can be found there exists a null vector $\mathrm{I}$, related to the static scalar fields, trivially 
\begin{align}
\label{null 1}
I^\m=:\int_{\emph{all\ space}}\ d^{d-1} x\ x^\m\p_0 T^{0\m}_{\mathrm{SFT}},
\end{align}
where integration can be done on the spatial parts of the space time and of course, for all values of $\m$, $I^\m\equiv0$.
This is because, due to the static
fields, the tensor of the energy-momentum is time independent.

Principally, with a superficial glance at classical electrodynamics, we know that the localized density of the charges
generates the decreasing fields faster than $r^{-\frac{3}{2}}$.
Now, let us have the following solution in which these fields
must tend to zero at infinite boundaries
faster than $\textbf{x}^{\frac{d-1}{2}}$.
For a positive number $\epsilon$ larger than $\frac{1}{2}$ the above can be written as
\begin{align}
\label{decrasing condition}
lim_{|\textbf{x}|\rightarrow\infty}|\textbf{x}|^{\frac{d-1}{2}+\epsilon}\phi(\textbf{x})\rightarrow0.
\end{align}
This implies that $\phi$, is a localized field and $T^{\m\n}_{\mathrm{SFT}}$, is a local conserved tensor.
By
employing part by part integration techniques
and exploiting the equation of the motion of the energy - momentum tensor,
Eq.(\ref{null 1}) is equivalent to the following
\begin{align}
I^i=-\int_{\emph{all\ space}}\ d^{d-1} x\ x^i\p_j T^{ji}_{\mathrm{SFT}},
\end{align}
with no summation of the index of $i$.
If we take into consideration that the fields decrease on the space boundaries, we will see that the first term vanishes;
therefore,
\begin{align}
I^i=\int_{\emph{all\ space}}\ d^{d-1} x\ T^{ii}_{\mathrm{SFT}}.
\end{align}
Then from Eq.(\ref{energy - momentum}) we calculate the trace of the tensor of the energy - momentum.
It is given by the following:
\begin{align}
T^i_{i\ \mathrm{SFT}}=-(d-1)\mathfrak{L}_{\mathrm{SFT}}-\p^i\phi\ \p_i \phi,
\end{align}
so we have
\begin{align}
\label{sigmanull}
I^i_{i\ \mathrm{SFT}}=\int_{\emph{all\ space}}\ d^{d-1} x\ ((1-d)
\mathfrak{L}_{\mathrm{SFT}}-\p^i\phi\ \p_i \phi)
\cr
=\int_{\emph{all\ space}}\ d^{d-1} x\ ((\frac{d-1}{2})(\p_i\phi\ \p^i\phi+m^2\phi^2)
-\p_i\phi\ \p^i\phi),
\end{align}
the last equation arises by employing the explicit definition of $\mathfrak{L}_{\mathrm{SFT}}$.
By using the equation of the motion of the field  and working with part by part integration techniques, we have
\begin{align}
\label{mass}
\int_{\emph{all\ space}}\ d^{d-1} x\ m^2\phi^2=0.
\end{align}
During the positive integrand function, the considered scalar fields
should be null in all Minkowski space time dimensions except for $d=2$, that is $\phi\equiv0$ besides two dimensional space time.
Always, it is possible to specific a dynamic Lump in order that the Lump
will be in rest because the free scalar Lumps are massive and there can always be found a Lorentz transformation.
So, we can generalize the lemma which consists of the time dependent solutions.
By proposing the stronger one:\\

\textbf{Stronger Lemma A}: \textit{The Lumps of scalar fields are forbidden in all Minkowski space time dimensions 
except in $1+1$ dimensional space time}.\\

Now, we continue extracting the Lump that consists of the pure interacting scalar fields.
We now want to determine the absence of interacting Lump in terms of the dimensional space time.
The interacting scalar field together with the pure self - interacting potential $V(\phi)$,
is given by
\begin{align}
\mathfrak{L}_{\emph{int}}=-\frac{1}{2}(\eta^{\m\n}\p_\m\phi\ \p_\n\phi+m^2\phi^2)-\frac{\lambda}{n!}\phi^n,
\end{align}
where $\lambda>0$ for the energy coast.
The equation of the motion of the scalar fields becomes
\begin{align}
\label{interacting motion equation}
\p_\m(\eta^{\m\n}\p_\n)\phi-m^2\phi-\frac{\lambda}{(n-1)!}\phi^{n-1}=0,
\end{align}
the tensor of the energy - momentum, $T^{\m\n}_{\mathrm{SFT}\emph{int}}$, which can be obtained from Eq.(\ref{energy - momentum})
is a conserved local tensor.
The null vector $I^\m$ is still available while the scalar field satisfies Eq.(\ref{decrasing condition}),
that is
\begin{align}
I^i=\int_{\emph{all\ space}}\ d^{d-1} x\ T^{ii}_{\mathrm{SFT}\emph{int}}=0.
\end{align}
Also,
we calculate the trace of the tensor of the energy - momentum in the following
\begin{align}
T^i_{i\ \mathrm{SFT}\emph{int}}=-(d-1)\mathfrak{L}_{\emph{int}}-\p^i\phi\ \p_i \phi.
\end{align}
Similar to Eq.(\ref{sigmanull}) we reach
\begin{align}
I^i_i=\int_{\emph{all\ space}}\ d^{d-1} x\ (\frac{(d-3)}{2}\p_i\phi\ \p^i\phi
+\frac{(d-1)}{2}m^2\phi^2
\cr
+\frac{(d-1)}{n!}\lambda \phi^n)=0,
\end{align}
again, if we employ the Eq.(\ref{interacting motion equation}), we reach
\begin{align}
\label{interacting condition}
I^i_i=\int_{\emph{all\ space}}\ d^{d-1} x\ (m^2\phi^2+\frac{\lambda}{2n!} f(n,d)\phi^n)=0,
\end{align}
where
\begin{align}
f(n,d)=3n-nd+2d-2.
\end{align}
For research on the conditions in the absence of the Lump, the sign of the second term 
must be same as the first term.
Obviously, this statement is not true for $f(2,d)$, because it is always positive and this is in contrast to the Derrick and Sine - Gordon theories.
Whereas, it can be seen that the obtained equation is in agreement with recalled theorems.
However, absence of the Lumps occurs at
\begin{align}
d\leqslant \frac{3n-2}{n-2}.
\end{align}
This says that the Lump can not occurs in a space time the dimensions of which would satisfy the above equation.
That is for $n=4$, we have no Lumps in $d\leqslant 4+1$, space time.
We learn nothing further from the above for the higher dimensions.
So, the free Lumps can occur in $d>1+\frac{2n}{n-2}$.
For the presence of the self-interaction of the scalar field with mixed potentials such as $\frac{\lambda_1}{4!}\phi^4+\frac{\lambda_2}{6!}\phi^6$
we reach a new condition.
Instance, the absence of interacting Lumps occurs in $d\leqslant 1+\frac{2\star6}{6-2}\leqslant3+1$ for the case of potential ($\frac{\lambda_1}{4!}\phi^4+\frac{\lambda_2}{6!}\phi^6$).
By applying the obtained condition for the Sine - Gordon theory, $V(\phi)\sim\cos{(\phi)}$, and setting 
$n\rightarrow\infty$, we reach to the absence dimensions of the Sine - Gordon solitons, $d\leqslant 2+1$.
But, from the previous section, we have found that the Sine - Gordon solitons can be occurred in $1+1$ space time dimensions.
So, the Sine - Gordon fields are absent only in space time with dimension $2+1$.
It is
possible to specify a dynamic interacting Lump in order that the interacting Lump
will be in a rest frame.
We say that a Lorentz transformation can be found, then we can generalize these lemmas which consist of the time dependent solutions.
So, the stronger lemma becomes\\

\textbf{Stronger Lemma B}: \textit{The dynamics Lumps of Lorentz invariance interacting potential of scalar fields can not occur
in the Minkowski space time at $d\leq 1+\frac{2n}{n-2}$ }
\textit{and greater interacting term in a given potential can always apply the final condition}.\\
Clearly, if $n$ gets the odd number then we learn nothing further from the Eq.(\ref{interacting condition}).
Similarly, for the Gluons \cite{coleman} and Proca cases, we can reach to the following equations:
\begin{align}\label{pfe}
\int_{\mathrm{all\ space}}d^{d-1} x\ T^i_{i\ \mathrm{Gluon}}
\cr
=\int_{\mathrm{all\ space}}d^{d-1} x\ (\frac{3-d}{2}f_{i0}^2+\frac{d-5}{4}f^2_{ij})=0,
\end{align}
and
\begin{align}\label{gfe}
\int_{\mathrm{all\ space}}d^{d-1} x\ T^i_{i\ \mathrm{Proca}}
\cr
=\int_{\mathrm{all\ space}}d^{d-1} x\ (\frac{3-d}{2}f_{i0}^2+\frac{d-5}{4}f^2_{ij}+\frac{1-d}{2}m^2A^2)=0.
\end{align}
Although, the compactness of the gauge group needs the first and second terms in Eqs.(\ref{pfe}-\ref{gfe}) have the same sign for $d=4$, 
so the classical Yang-Mills theory in four-dimensional
Minkowski space has no Lumps wherase the Lumps of Proca can occur.

\section{Generalization of The Theory}

In this section the formulation is specialized to the case for 
scalar field.
The equivalence between the Minkowski space time and the local pseudo Riemann's manifolds,
implies that
the forbiddance of the scalar fields in the Minkowski space time can be generalized to the general universe consisting of gravitational backgrounds.
There are two viewpoints to be studied in related to perturbations theoretical physics in the presence of background
gravitational effects.
The first, is rewriting quantum mechanics on a curved background
space time exclusively, named DeWitt's method.
The second, was introduced by
Weber as he was concerned with the response of matter to the presence of
the gravitational effects.
These approaches are not equivalent and any results
obtained from one will no necessarily be obtained from the other\cite{speliotopoulos,saha}.
In this study, the generalization of the absence of the scalar fields is based on Weber's interpretation and
there is no warranty of extracting similar results through Dirac's method.
The classical Lagrangian density for the scalar field in a general space time is given by\cite{inverno, kleinert1, kleinert2, wald, birrell}
\begin{align}
\label{ordinary action}
\mathfrak{L}=-\frac{1}{2}\sqrt{-g}(g^{\m\n}\p_\m\phi(x)\p_\n\phi(x)+(m^2+\xi R(x))\phi^2),
\end{align}
where $m$ is the bare mass of the scalar field corrected by the scalar curvature, $R(x)$, consistent
with the dimension of space time $(\xi=\frac{(d-2)}{4(d-1)})$ and
$g^{\m\n}$ is the metric tensor of general space time
which is different from the Minkowski metric given in the last section.
According to Eqs.(\ref{metric}-\ref{affine}) the first approximation makes $R(x)$ vanish.
The operator $\nabla_\m$ is a covariant derivative which for the scalar field will be
$\nabla_\m\phi=\p_\m\phi$.
So, the equation of the motion and the Laplace Beltrami operator become
\begin{align}
\label{equation of motion} &&
(\Box-m^2-\xi R)\phi=0,
\end{align}
where $\Box=\frac{1}{\sqrt{-g}}\p_\m(\sqrt{-g}g^{\m\n}\p_\n)$.
In the above analysis, the equation of the motion of the nonrelativistic fields get the Christoffel symbols 
without also including the minimal coupling term.
The vanishing of the minimal coupling term 
arises from the trivial null Ricci's tensors.
From the Eqs.(\ref{metric}-\ref{affine}), the first approximation makes to vanishing of the $R(x)$
\cite{parker1, kleinert2, fulling, wald, birrell}.
In the DeWitt method, the starting point for the investigation of a typical interacting problem(such as the Hydrogen atom) is based on
fundamental theory. 
In contrast, in the Weber method, we must make an effective theory from the approximated equation of motion.
The approximations
should be done on Eqs.(\ref{equation of motion}).
Another part of the approximation consists of changes the Laplace Beltrami operator.

In our scenario, the equation of the motion of the scalar field should be considered in the tangent space.
By applying the geometrical effects on the equation of the motion, the compatible theory is derived.
The geometrical effects appear in the Laplacian operator.
Finally, they appear in
the equation of the motion
\begin{align}
\label{weber lagrangy}
(\p_\m\p^\m+\Gamma^\m_{\m\a}\p^\a-m^2)\phi=0.
\end{align}
The geometrical effects are induced by approximating Christoffel symbols via Eq.(\ref{affine}).
Additional approximations exist according to the tangent space consideration vanishing Ricci scalar which is related to the
ignorance of the direct interaction between
matter and geometry. It obviously breaks the conformal invariance of the theory.
The second term is a new interaction nonzero Christoffel which appears allying the derivation of the fields and gravity.
It is easy to see that, according to Eqs.(\ref{affine}), $\Gamma^\m_{\m0}=0$, and for concentrate, we can introduce
\begin{align}
\tilde{\Gamma}_{ij}=R_{0i0j}+\frac{1}{3}R_{jiij},
\end{align}
with these tools, the gravitational effects appear in the Eq.(\ref{weber lagrangy}) get the form: $\mathbf{x}\cdot\tilde{\mathbf{\Gamma}}\cdot\mathbf{\nabla}\phi$,
so we can suggest 
the following Lagrangian which is given on the BCK model \cite{bck1, bck2, bck3}
can satisfy Eq.(\ref{weber lagrangy})
\begin{align}
\label{damping lagrangy}
\mathfrak{L}_{_\emph{web}}=e^{\frac{1}{2}\mathbf{x}\cdot\tilde{\mathbf{\Gamma}}\cdot\mathbf{x}}(-\frac{1}{2}\eta^{\m\n}\p_\m\phi\ \p_\n\phi-\frac{1}{2}m^2\phi\phi)
=
e^{\frac{1}{2}\mathbf{x}\cdot\tilde{\mathbf{\Gamma}}\cdot\mathbf{x}}\mathfrak{L}_0,
\end{align}
where $\mathfrak{L}_0=-\frac{1}{2}\eta^{\m\n}\p_\m\phi\ \p_\n\phi-\frac{1}{2}m^2\phi\phi$.
The factor($\frac{1}{2}$) which appears in front of the exponential function is the controller of the symmetric structures
of the static Schwarzschild metric.
For the static Schwarzschild universe with spherical symmetry, we have
\begin{align}
ds^2=-(1-\frac{2Gm}{r})dt^2+(1-\frac{2Gm}{r})^{-1}dr^2+r^2(d\t^2
\cr
+\sin^2{\t}d\phi^2).
\end{align}
The suggested tangent space is perpendicular to the world line and is constructed
on geodesics coordinates.
The non vanishing $R_{\m\n\a\b}$ are the relevant spatial components of Riemann's tensors
for the spherical symmetry of Schwarzschild metric given by
\begin{align}
R^1_{010}=\frac{2GM(2GM-r)}{r^4},R^2_{020}=\frac{GM(-2GM+r)}{r^4},
\cr
R^3_{030}=\frac{GM(-2GM+r)}{r^4},R^3_{232}=\frac{2GM}{r},
\cr
R^3_{131}=\frac{GM}{r^2(2GM-r)},R^2_{121}=\frac{GM}{r^2(2GM-r)}.
\end{align}
For the considered tangent space, it can be seen that the components of Riemann's tensors will be constants such as
$R^1_{010}\approx\frac{-2GM}{R^3}$ where $R$ is the large distance of center of mass the assumed quantum system from the 
massive body which we considered.
It is easy to show that the Lagrangian Eq.(\ref{damping lagrangy}) satisfies the equation of
the motion Eq.(\ref{weber lagrangy}).
The front factor of the Lagrangian Eq.(\ref{damping lagrangy}) can be absorbed directly by $\phi$,
\begin{align}
\label{cannonical transformation}
\exp(\frac{1}{4}\mathbf{x}\cdot\tilde{\mathbf{\Gamma}}\cdot\mathbf{x})\phi=\acute{\phi},
\end{align}
so, we have
\begin{align}
\mathfrak{\acute{L}}_{_\emph{web}}=
-\frac{1}{2}\p_\m\acute{\phi}\ \p^\m\acute{\phi}-\frac{1}{2}\tilde{\Gamma}_{\m j}x^j
\acute{\phi}\ \p^\m\acute{\phi}-\frac{1}{2}m^2 \acute{\phi}\acute{\phi},
\end{align}
where, $\tilde{\Gamma}_{0j}=0$.
If we ignore the total derivations; therefore, $\mathfrak{\acute{L}}_{_\emph{web}}$ becomes
\begin{align}
\label{important}
\mathfrak{\acute{L}}_{_\emph{web}}=
-\frac{1}{2}(\p_\m\acute{\phi}\ \p^\m\acute{\phi}+(m^2-\frac{1}{2}\Sigma_j\tilde{\Gamma}_{jj})\acute{\phi}\acute{\phi})
\cr
\equiv
\mathfrak{L}_0(\acute{\phi},\p_\m\acute{\phi}\ ;\ m^2\rightarrow m^2-\frac{1}{2}\Sigma_j\tilde{\Gamma}_{jj}).
\end{align}
It confirms the approximately solutions Eq.(\ref{equation of motion})
also a new mass shell is given by $m^2-\frac{1}{2}\Sigma_j\tilde{\Gamma}_{jj}$.
Although, the energy - momentum of 
the generalized scalar field that appears in the Eq.(\ref{weber lagrangy}) is not locally conserved by itself, ($\p_\m T^{\m\n}\neq0$).
But, 
it can be shown that there is an undefined locally conserved current, 
($\p_\m \acute{T}_{\emph{web}}^{\m\n}(\acute{\phi},\acute{m})=0$).
So, locally, we reach to the standard scalar field theory with a slight change in mass
where
\begin{align}
T^{\m\n}_{_{\emph{web}}}\sim T^{\m\n}_{0}(\phi\rightarrow\acute{\phi},m^2\rightarrow m^2-\frac{1}{2}\Sigma_j\tilde{\Gamma}_{jj}),
\end{align}
and this is given by Eq.(\ref{energy - momentum}).
Although, according to the BCK model and extracting mirror fields,
it can be shown that there is no conserved tensor of energy momentum of the scalar fields in general space time,
but, the following is available
\begin{align}
\label{new field}
\int_{\emph{all\ space}}\ d^{d-1} x\ \acute{m}^2e^{\frac{1}{2}\mathbf{x}\cdot\tilde{\mathbf{\Gamma}}\cdot\mathbf{x}}\phi^2=0,
\end{align}
therefore, all the above is still valid for the Lump in the tangent spaces.
Now, we can say that if the Lump can not occur locally , it is absent
globally because of the Homomorphism between the Minkowski and pseudo
Riemann's manifolds.
Of course, there is no vice versa state.
This means that if the result of local research is be empty,
this shall also be the case globally.
So we can propose the following universal theorems
From the Eq.(\ref{new field}), the fields must vanish at infinity faster than $r^{\frac{d+3}{2}}$ for the absence of the Lumps.
We mention that the factor of $e^{\frac{1}{2}\mathbf{x}\cdot\tilde{\mathbf{\Gamma}}\cdot\mathbf{x}}$ that appears in the Eq.(\ref{new field}) has 
a good interpretation, because we need to extend our explicit formula up to the first order of the Christoffel symbols.
However, the previous lemmas, consists interacting Lumps, can not be generalized comprehensively.
So, we can propose the following universal theorems,

\textbf{Lemma 1}: \textit{In the Weber's method, nonrelativistic scalar field is similar to the case of the Minkowski with replacing
mass with deviation.}

\textbf{Lemma 2}: \textit{The pure interacting Lump is forbidden
in general space time where the general space time dimensions satisfy the degree of potential: $d\leq 1+\frac{2n}{n-2}$
which the Lumps must go to zero at infinity faster than $r^{\frac{d+3}{2}}$}.

\section{Discussion}

This paper is based on the dynamics of the fields and the energy - momentum tensor. 
From the dynamics of the energy - momentum viewpoint, we show that the classical Lumps are absent for all values of $d$-dimensional space time except
$d=2$.
Also, we show that the absence of the interacting relevant scalar Lump is directly related to the interacting potential.
Due to Weber's method, we can rewrite the scalar field in the presence of the gravitational backgrounds
and according to the BCK model which describes the damping oscillator, we can suggest a new Lagrangian for the nonlinear equation of the
motion of the scalar field
and also
by exploiting the Homomorphy between the Minkowski and pseudo Riemann's manifolds, we can generalize the absence of Lumps onto general space time.
\section{Acknowledgments}
The author thanks the Shahrekord University for support of this research.
\newline



\begin{thebibliography}{99}

\bibitem{derrick} G.H. Derrick, J. Math. Phys. {\bf 5} (1964) 1252.

\bibitem{kolokolov} N.G. Vakhitov and A.A. Kolokolov, Radiophys. Quantum Electron. {\bf 16} (1973) 783.

\bibitem{coleman1} Commun. math. Phys. {\bf 55} (1977) 113.

\bibitem{deser} S. Deser, Phys. Lett. {\bf B64} (1976) 463.

\bibitem{jafari} A.H. Fatollahi, A. Jafari, Eur. Phys. J. {\bf C46} (2006) 235.

\bibitem{coleman2} Commun. math. Phys.  {\bf 31} (1973) 259.

\bibitem{speliotopoulos} A.D. Speliotopoulos, Phys. Rev. {\bf D51} (1995) 1701.

\bibitem{saha} A. Saha, S. Gangopadhyay, S. Saha, Phys. Rev. {\bf D83} (2011) 025004.

\bibitem{raja} R. Rajaraman, "Solitons and Instantons: An Introduction to Solitons and Instantons in Quantum Field Theory", (North-Holland Personal Library) (1989). 

\bibitem{dod} K. R. Dodd, J. C. Eilbeck, J. D. Gibbon, H. C. Morris, x"Solitons and Nonlinear Wave Equations", (Academic Press. London) 1989.

\bibitem{nestrov} A. I. Nestrov, Class. Quant. Grav. {\bf 16} (1999) 465-477.

\bibitem{misner} C.W. Misner, K.S. Thorne, J.A. Wheeler, "Gravitation", (Freeman Publishing Company, San Francisco) 1973.

\bibitem{weber} J. Weber, "General Relativity and Gravitational Waves", (Interscience Publisher INC, New York,
Dover Edition) 2004.

\bibitem{maggiore} M. Maggiore, "Gravitational Waves", (Oxford University Press INC, New York) 2008.

\bibitem{inverno} R. D'Inverno, "Introducing Einstein's Relativity", (Oxford University Press Inc, New York) 1993.

\bibitem{parker} L.E. Parker, Phys. Rev. {\bf D22} (1980) 1922.

\bibitem{parker1} L.E. Parker, D.J. Toms, "Quantum Field Theory on curved Space Time", (Cambridge University Press, New York) 2009.

\bibitem{kleinert1} H. Kleinert, "Path integral in Quantum Mechanics, Statistics,
Polymer Physics, and Financial Markets", (World Scientific Publishing Company) 2009.

\bibitem{kleinert2} H. Kleinert, "Multivalued Fields: In Condensed Matter, Electromagnetism and Gravitation
", (World Scientific Publishing Company) 2008.

\bibitem{coleman} S. Coleman, Commun. math. Phys. {\bf 55} (1997) 113-116.

\bibitem{wald} R. M. Wald, "Quantum Field Theory in Curved Space Time and Black Hole Thermodynamics", (The University of Chicago Press, Chicago) 1994.

\bibitem{birrell} N.D. Birrell, P.C.W. Davies, "Quantum Fields in Curved Space", (Cambridge University Press) 1982.

\bibitem{fulling} S.A. Fulling, "Aspects of Quantum Field Theory in Curved Space Time", (Cambridge University Press) 1989.

\bibitem{bck1}H. Bateman, Phys. Rev. {\bf 38} (1931) 815.

\bibitem{bck2}P. Caldirola, Nuovo Cimento {\bf 18} (1941) 393.

\bibitem{bck3}E. Kanai, Prog. Theoret. Phys. {\bf 3} (1948) 440.

\end{thebibliography}
\end{document}